\documentclass[journal,transmag]{IEEEtran}

\usepackage{cite}

\ifCLASSINFOpdf
  \usepackage[pdftex]{graphicx}
\else
\fi

\usepackage{amsmath}
\usepackage{amssymb}
\usepackage{stfloats}
\usepackage{url}
\usepackage{xcolor}
\begin{document}
\title{Dynamics of a Spin-Wave Active Ring Resonator Driven by Harmonic-Null Square-Wave and Unipolar 8-bit Walsh Code Modulations}
\author{\IEEEauthorblockN{Anirban Mukhopadhyay\IEEEauthorrefmark{1},
Kaustubh Narayan\IEEEauthorrefmark{1},
Anil Prabhakar\IEEEauthorrefmark{1}}
\IEEEauthorblockA{\IEEEauthorrefmark{1} Department of Electrical Engineering,
Indian Institute of Technology Madras, Chennai 600036, India}% <-this % stops an unwanted space
\thanks{
% Manuscript received "MM DD", 2026; revised "MM DD", 2015. 
Corresponding author: Anirban Mukhopadhyay (email: anirban.ds.research@outlook.com).}}
%%%%%%%%%%%%%%%%%%%%%%%%%%%%%%%%%%
%%%%%%%%%%%%%%%%%%%%%%%%%%%%%%%%%%
\IEEEtitleabstractindextext{%
\begin{abstract}
Spin-wave active ring resonators (SWARRs) based on yttrium iron garnet (YIG) films exhibit rich nonlinear dynamics that make them promising platforms for physical reservoir computing. We present systematic and experimentally simple methods to characterize a SWARR's nonlinear behavior and memory. We first use a third harmonic elimination method to probe the nonlinear response. A drive frequency $f_\mathrm{d}$ is modulated by a square-wave pattern engineered to have a spectral null at $3/T$, which is then applied as 
input to the SWARR. The power spectra at the output of the YIG delay line allow us to identify five distinct regions within a drive frequency range of $2.15 < f_\text{d} <  2.2\ \text{GHz}$ where nonlinearity was observed as frequency peaks at $f_\mathrm{d} \pm \frac{3}{T}$. The STM duration of the SWARR was estimated to be approximately 300~ns using a modulation pattern derived from the sequency-ordered 8-bit unipolar Walsh family. The nonlinear dynamics of the SWARR were further quantified by decomposing its temporal response to analog Walsh pulses 
in terms of the input Walsh codewords. The proposed methods of harmonic elimination and Walsh-function decomposition together provide a practical and general framework for the design and optimization of tunable spin-wave reservoir computers.
\end{abstract}
%%%%%%%%%%%%%%%%%%%%%%%%%%%%%%%%%%
\begin{IEEEkeywords}
Magnonics, nonlinear magnetics, spin-waves (SWs)
\end{IEEEkeywords}}
%%%%%%%%%%%%%%%%%%%%%%%%%%%%%%%%%%
%%%%%%%%%%%%%%%%%%%%%%%%%%%%%%%%%%
\maketitle
\IEEEdisplaynontitleabstractindextext
\IEEEpeerreviewmaketitle
\section{Introduction}
\IEEEPARstart{A}{}spin-wave active ring resonator (SWARR) is a hybrid resonant system composed of a spin-wave delay line and a variable-gain unit, connected in a closed magneto-electronic feedback loop. The resonator eigenmodes are excited when the round-trip phase delay is an integer multiple of $2\pi$ \cite{WU2010163}. The resonance condition is governed by the group velocity of the spin-waves, which in turn is influenced by several parameters such as the magnetic film thickness, its saturation magnetization, the spin-wave propagation loss, the applied magnetic field, and the specific experimental configuration that excites either forward, backward, or surface spin-waves~\cite{prabhakar2009spin}.
 
Active ring circuits with multiple YIG delay lines have been used to realize combinatorial logic and memory devices~\cite{khitun2022comb, 
balynsky2023coop, Balinskyy2024}. The spin-wave active ring system has further served as a base for time-delayed reservoir computing (TDRC), leveraging its intrinsic nonlinear dynamics and the propagation delay of spin-waves (SWs)~\cite{watt2021, watt2021_1, Watt2023, Ustinov2024}. Recently, a numerical reservoir computing model based on the three-wave decay process of magnetostatic surface spin-waves (MSSWs) in a SWARR was used to denoise a Lorenz96 chaotic time series~\cite{Kostylev2025}. In these works, the nonlinearity and the memory properties of the SWARR-based reservoir were estimated via a parity-check (PC) and a short-term memory (STM) task, respectively \cite{furuta2018macromagnetic, jaeger2002tutorial}. 

It is well established that in nonlinear systems, harmonics and intermodulation frequency components absent in the drive signal arise as a consequence of the system's inherent nonlinearity. This principle has been exploited to study nonlinear mechanisms in superconductors and ferroelectric PZT capacitors by measuring higher-order harmonic and intermodulation distortion (IMD) products \cite{Cifariello2006, Andreone2007, Tai2012, Vasudevan2013}. Harmonic elimination has also been extensively used in analyzing the nonlinear transition shifts in magnetic recording~\cite{Valcu13}.
In our experiment, we inject a drive signal of frequency 
$f_\mathrm{d}$ modulated by a square wave pattern of period $T$. The 
duty cycle of the modulation pattern is chosen to place a spectral null at the $k^\mathrm{th}$ harmonic 
of the modulation frequency $\frac{1}{T}$. The spectral measurement at $f_\mathrm{d} \pm \frac{k}{T}$ quantifies the nonlinearity. This approach provides a systematic means of tuning the ring gain $G$ and drive frequency $f_\mathrm{d}$ to control the nonlinear behavior of the SWARR.
In a separate experiment, analog pulses derived from the Walsh codewords are applied as modulation inputs to the SWARR. The resulting temporal response of the 
SWARR is then captured and subsequently decomposed in terms of the input 
Walsh analog pulses provide an alternative systematic means of quantifying the 
nonlinear behavior of the system. We also estimated the duration of short-term memory in the SWARR. To achieve this, we use two modulation signals, derived from the two highest-sequency codewords of an 8-bit unipolar Walsh family. The short-term memory is quantified using the $L^2$-norm ratio. We observe our SWARR has an STM duration of approximately 300~ns.
%%%%%%%%%%%%%%%%%%%%%%%%%%%%%%%%%%
%%%%%%%%%%%%%%%%%%%%%%%%%%%%%%%%%%
\section{Experimental setup}
\begin{figure}[htbp]
\centering
\includegraphics[width=\columnwidth]{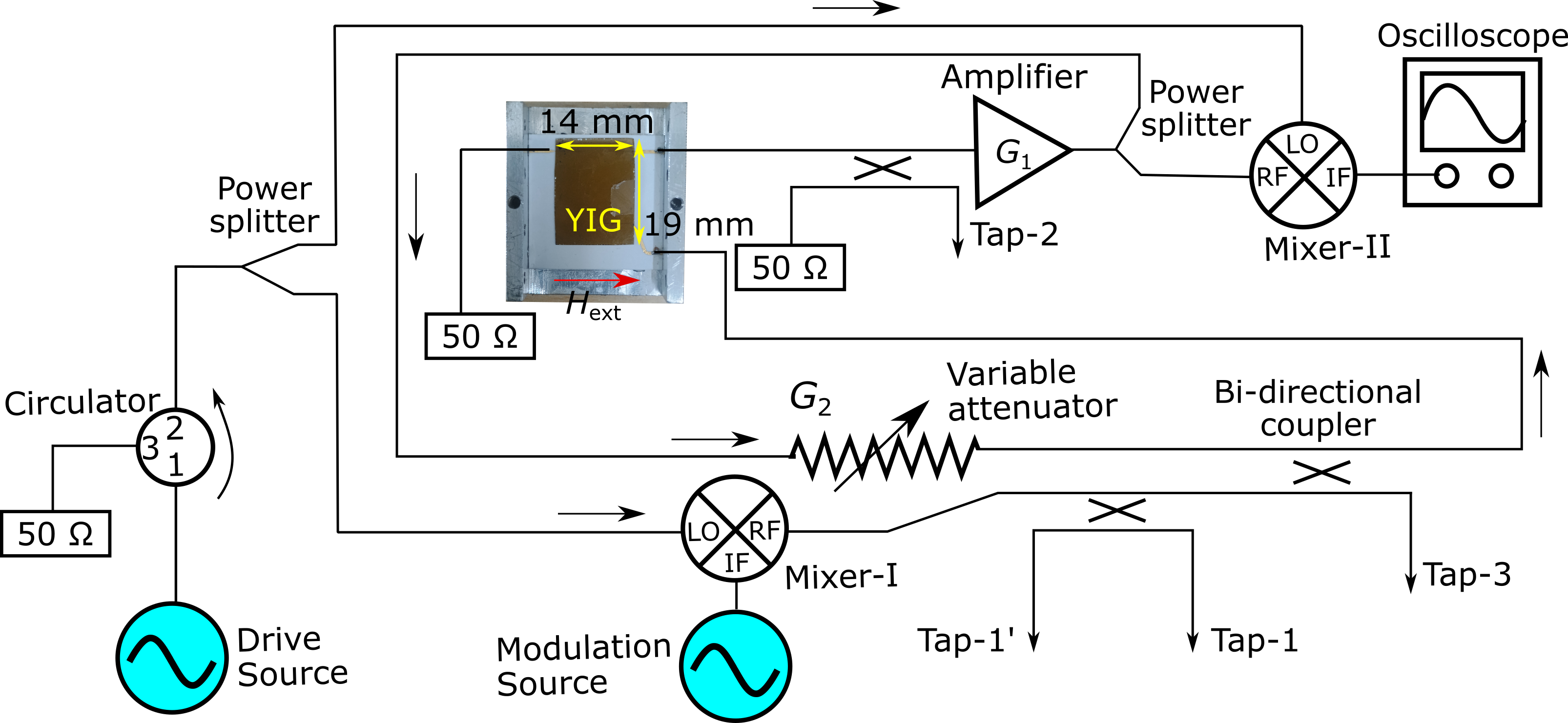}
\caption{Circuit diagram of a spin-wave active ring resonator (SWARR). Directional couplers were used to tap and measure the power spectrum at different locations in the resonator.}
\label{fig:swarr_ckt}
\end{figure}

A YIG film with a thickness of 6.9~$\mu$m serves as the spin-wave delay line. Together with a variable-gain unit, it forms the SWARR, shown in Fig.~\ref{fig:swarr_ckt}. The gain unit consists of a constant gain block amplifier $G_1 \approx 42~\mathrm{dB}$ and a variable attenuator $G_2$. The YIG film was saturated by an in-plane bias magnetic field applied perpendicular to the spin-wave propagation direction, such that the SWARR operated in the MSSW configuration. The film sits on top of two microstrip-line antennas. A bidirectional coupler injected the drive signals into the SWARR, while a power splitter extracted the output power. Two mixers were used to modulate the GHz drive signal and to demodulate the signal from the SWARR. The taps shown in Fig.~\ref{fig:swarr_ckt} were used to measure the power spectrum. The drive frequency $f_\mathrm{d}$ was varied from 2.14 to 2.195~GHz with a step size of 50~kHz. The spectrum analyzer's observation window $f_\mathrm{obs}$ was set from 2.12 to 2.2~GHz, with a RBW of 10~kHz. The SWARR operated at three different ring gains, $G = G_1 + G_2 \approx 33, 36$, and 39~dB. For every combination $(G, f_\mathrm{d})$, we captured the output power spectrum $P_\mathrm{out}(G, f_\mathrm{d}, f_\mathrm{obs})$ at Tap-2 (from the output of the MSSW delay line) and demodulated signal $y(G, f_\mathrm{d}, t)$ from the output of the mixer-II.

%%%%%%%%%%%%%%%%%%%%%%%%%%%%%%%%%%
\subsection{Designing periodic modulation square wave}
To design a periodic square-wave modulation signal \mbox{$x(t+T) = x(t)$} without the $k^\mathrm{th}$ harmonic component, one has to set the duty-cycle $D$ according to (3.44) in \cite{Oppenheim1996}:
\begin{align}
    k D &= m,\:\:m \in \mathbb{N}_1 = \{1, 2, 3,\ldots\}.
\end{align}
A duty cycle of $D = \frac{1}{3}$ was chosen, which suppressed the third harmonic and all higher harmonics that were integer multiples of $\frac{3}{T}$. $T$ was fixed at 1~$\mu$s, and the modulated drive signal had frequency components $f_\mathrm{d} \pm \frac{i}{T}$, where $i \neq 3m$.
Spectral response at $f_\mathrm{d} \pm \frac{3}{T}$ acts as a metric for nonlinearity in SWARR.
%%%%%%%%%%%%%%%%%%%%%%%%%%%%%%%%%%
\subsection{Creating modulation signals from unipolar Walsh family}
\begin{figure}[!htbp]
\centering
\includegraphics[width=0.85\columnwidth]{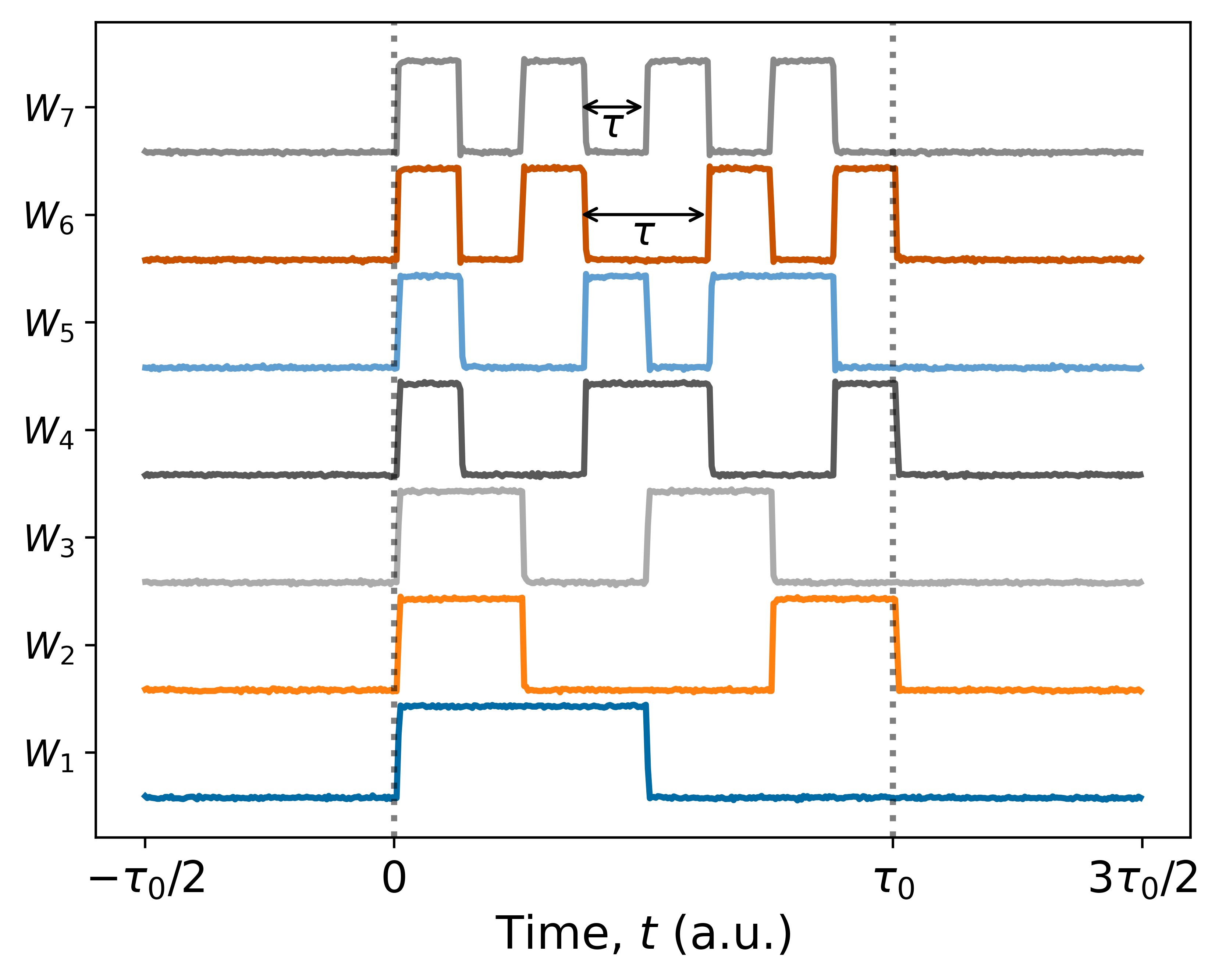}
\caption{Analog pulses generated from discrete Walsh functions $\mathbf{\tilde{W}}$. The pair of vertical dashed lines indicates duration $\tau_0$.}
\label{fig:w_67}
\end{figure}
For a code of length $N = 2^m$, $m \in \mathbb{N}_1$, the bipolar Walsh matrix $\mathbb{W}_N$ is defined as follows:
\begin{equation}
    \mathbb{W}_{2^m} = \begin{bmatrix} \mathbb{W}_{2^{m-1}} & \mathbb{W}_{2^{m-1}} \\ 
    \mathbb{W}_{2^{m-1}} & -\mathbb{W}_{2^{m-1}} \end{bmatrix}, \quad
     \mathbb{W}_{2} = \begin{bmatrix} 1 & 1 \\ 
    1 & -1 \end{bmatrix}.
    \label{eq:bipolarwalsh}
\end{equation}
To construct sequency-ordered $N$-bit Walsh codes, the rows $\mathbb{W}_{2^m}$ are arranged in order of 
increasing sequency, i.e., the number of zero-crossings, analogous to frequency in sinusoidal signals~\cite{walsh1923closed, beauchamp1976walsh}.
The unipolar Walsh matrix $\mathbf{W}_{2^m}$ can be derived from
(\ref{eq:bipolarwalsh}),
\begin{equation}
    \mathbf{W}_{2^m} = \frac{\mathbb{W}_{2^m} + \mathbf{J}_{2^m}}{2},
    \label{eq:unipolarwalsh}
\end{equation}
where $(\mathbf{J}_{2^m})_{ij} = 1\ \forall\ i,j$ is the all-ones matrix.

Analog modulation pulses were generated by applying a zero-order hold 
(ZOH) operation to each bit of the 8-bit unipolar Walsh family, excluding 
the zeroth-order (all-ones) row, such that $\mathbf{\tilde{W}}_8(t) = [W_1, W_2, \ldots, W_7]$. These input pulses are applied to the SWARR, and its temporal response to each input is recorded.
\begin{equation}
    \mathbf{\tilde{Y}}_8(f_\mathrm{d}, t) = \begin{bmatrix}
        y_1\\
        \vdots\\
        y_7
    \end{bmatrix} \approx \begin{bmatrix}
        c_{10} & \ldots & c_{17}\\
        \vdots & \ddots & \vdots \\
        c_{70} & \cdots & c_{77} 
    \end{bmatrix} 
    \begin{bmatrix}
        W_0\\
        \vdots\\
        W_7
    \end{bmatrix} = \mathbf{C}(f_\mathrm{d}) \mathbf{W}_8(t),  
\end{equation}
where $W_i$ and $y_i$ are the input Walsh pulse and its corresponding response from SWARR, respectively. The root mean square error (RMSE) between the actual signal and its reconstruction is defined as,
\begin{equation}
    \mathbf{E}(f_\mathrm{d}) = \begin{bmatrix}
        e_1\\
        \vdots\\
        e_7
    \end{bmatrix} = 
        \sqrt{\langle \left(\mathbf{\tilde{Y}}_8 - \mathbf{C} \mathbf{W}_8\right)^2 \rangle_t},
\end{equation}
where $\langle \cdot \rangle_t$ indicates temporal averaging.
A higher value of $e_i$ indicates greater shape deformation of the 
response, and hence stronger nonlinearity.
The STM duration was estimated using the two highest-sequency codewords $W_{6,7}(t)$, expressed as,
\begin{equation}
    W_i(t) = \sum_{k=0}^{7} b_k^{(i)} \cdot 
    \mathrm{rect}\!\left(\frac{t - \left(k + \frac{1}{2}\right)
    \frac{\tau_0}{8}}{\frac{\tau_0}{8}}\right), \quad i \in \{6,7\},
    \label{eq:walsh_analog}
\end{equation}
where, $\tau_0$ is the code duration, $b_j^{(i)}$ denotes the $j^\mathrm{th}$ bit of codeword $W_i$. $\left[b_0^{(6)}, b_1^{(6)}, \ldots, b_7^{(6)}\right] = 
[1,0,1,0,0,1,0,1]$ and $\left[b_0^{(7)}, b_1^{(7)}, \ldots, 
b_7^{(7)}\right] = [1,0,1,0,1,0,1,0]$.
In our experiments, the analog modulation patterns for $W_{6,7}(t)$ had code durations of $\tau_0 = 1$, 1.2, and 1.6~$\mu$s, and the peak modulation amplitude was 100~mV. We have marked in Fig.~\ref{fig:w_67} the distance $\tau$ that exhibits varying time-length for different $\tau_0$ and is therefore well-suited for probing temporal memory.
%%%%%%%%%%%%%%%%%%%%%%%%%%%%%%%%%%
%%%%%%%%%%%%%%%%%%%%%%%%%%%%%%%%%%
\section{Experimental results}
\begin{figure*}[!htbp]
\centering
\includegraphics[width=0.7\textwidth]{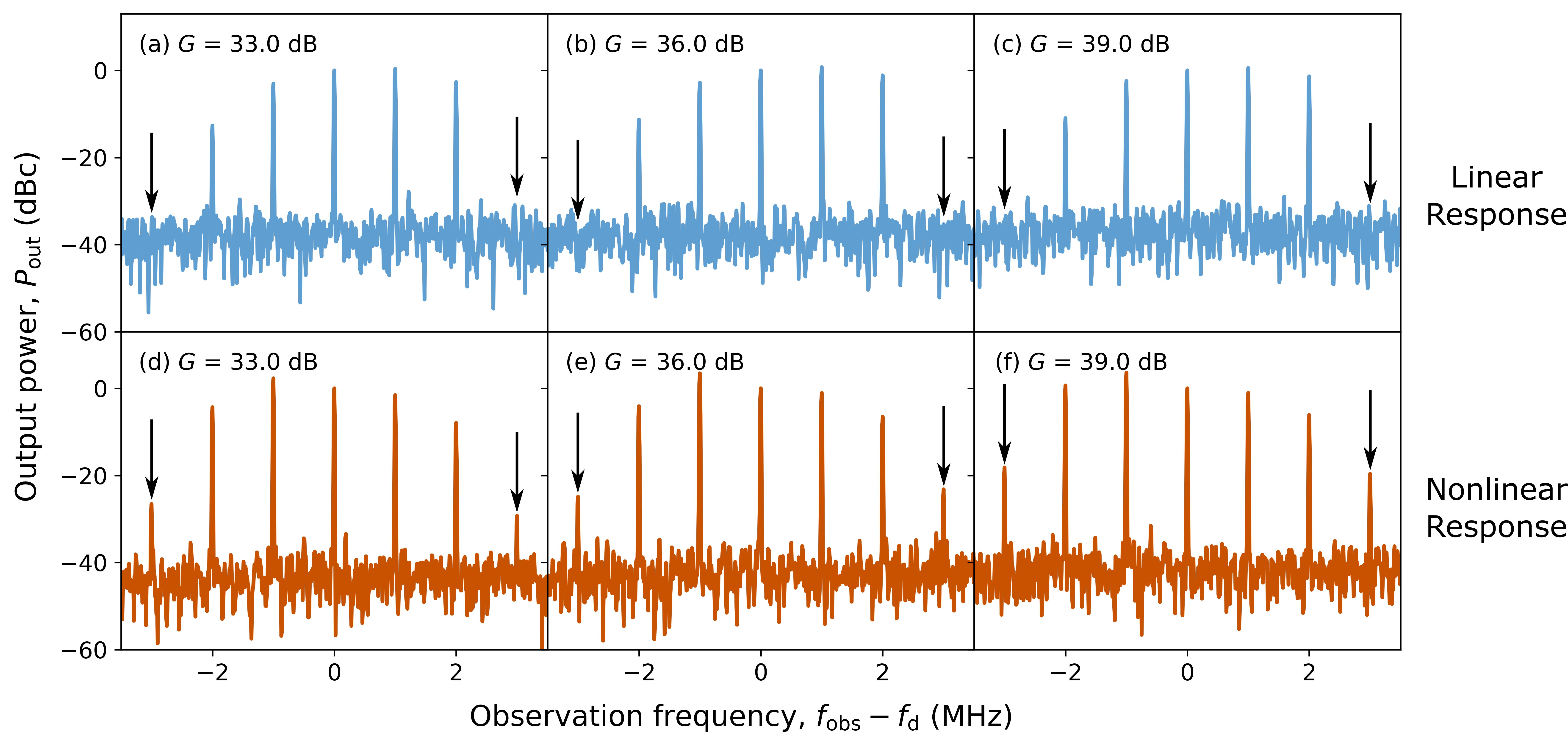}
\caption{Power spectra extracted from Tap-2 of the SWARR driven at (a-c) 2.142~GHz and (d-f) 2.1588~GHz. In (a-c), the spectral components at $f_\mathrm{d} \pm \frac{3}{T}$ are absent, whereas in (d-f), a gradual increase in power at these frequencies is observed with increasing gain $G$. The presence of the third harmonic indicates a strong nonlinear response.}
\label{fig:spectra}
\end{figure*}
\subsection{Nonlinearity with 3$^{rd}$ harmonic elimination patterns}
\begin{figure}[!htbp]
\centering
\includegraphics[width=0.85\columnwidth]{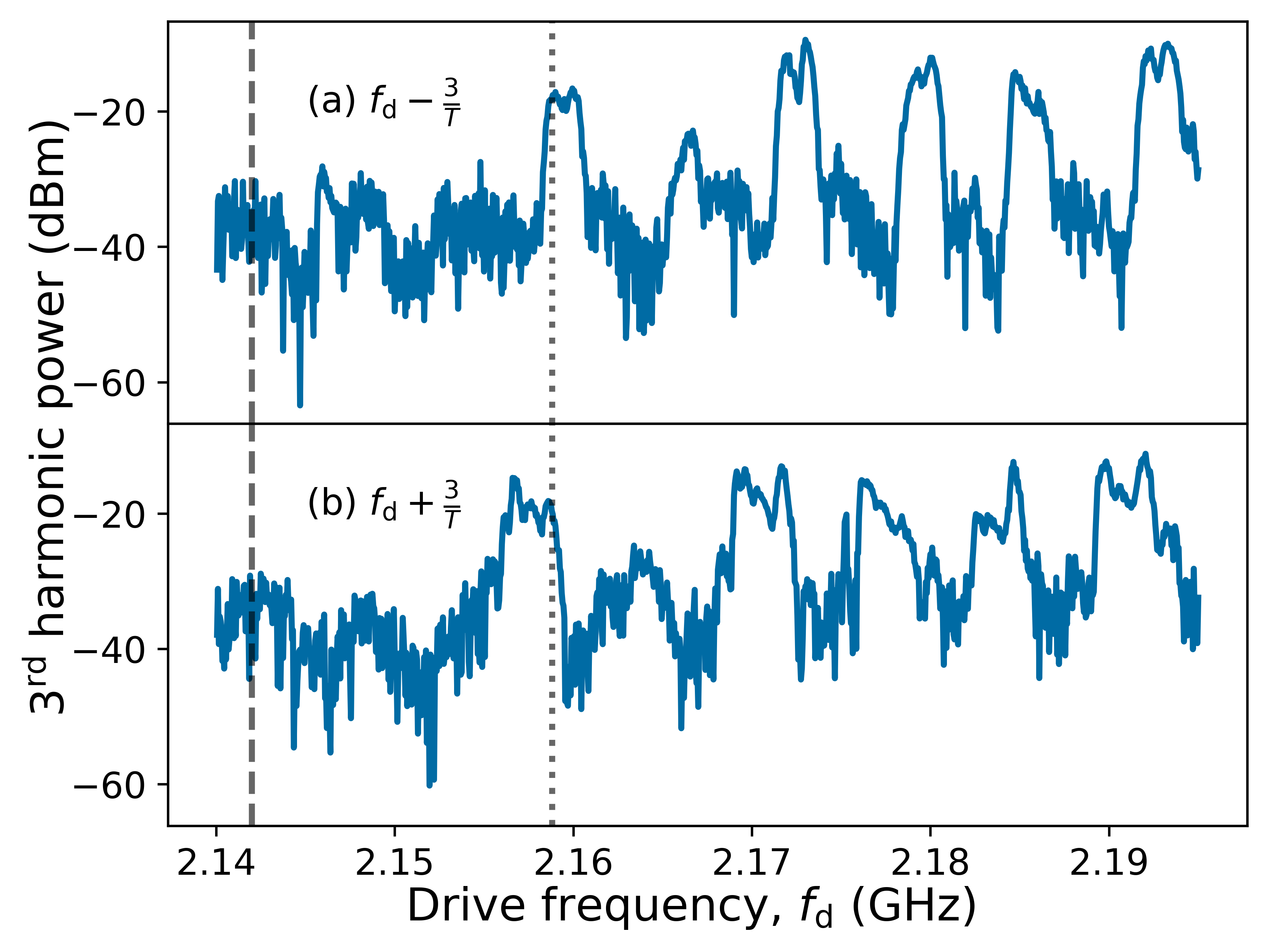}
\caption{Output power $P_\mathrm{out}$ at $f_\mathrm{d} \pm \frac{3}{T}$ as function of $f_\text{d}$ at $G = 39~\mathrm{dB}$. The dashed and dotted vertical lines indicate $f_\mathrm{d} = 2.142$ and 2.1588~GHz.}
\label{fig:pwr_3f}
\end{figure}

We captured the spectrum at Tap-2, shown in Fig.~\ref{fig:swarr_ckt}, at each $f_\mathrm{d}$. 
We have extracted $P_\mathrm{out}(G\approx39~\mathrm{dB}, f_\mathrm{d}, f_\mathrm{d}\pm\frac{3}{T})$ from these captured spectra and plot them in Fig.~\ref{fig:pwr_3f}. We are able to identify five regions in the power spectrum, as we sweep the drive frequency, where the third harmonic is excited. We also note that the nonlinear bands corresponding to $f_\mathrm{d}-\frac{3}{T}$ are shifted to higher frequencies by approximately 2~MHz compared to that for $f_\mathrm{d}+\frac{3}{T}$ at $G \approx 39~\mathrm{dB}$. This could be related to the opening of the spin wave cone angle or just a thermal effect, and it deserves further investigation. But for now, we focus on identifying the nonlinear regions.
\subsection{Spectral output from Tap-2 of SWARR}
Fig.~\ref{fig:spectra} shows power spectra extracted from Tap-2 for $f_\mathrm{d} = 2.142$ and 2.1588~GHz across all three $G$ values. The third harmonic is absent at $f_\mathrm{d} = 2.142~\mathrm{GHz}$, while it emerges and grows with gain at $f_\mathrm{d} = 2.1588~\mathrm{GHz}$, consistent with the nonlinear response that we see in Fig.~\ref{fig:pwr_3f}.
%%%%%%%%%%%%%%%%%%%%%%%%%%%%%%%%%%
\subsection{Nonlinearity with Walsh codes}
\begin{figure}[!htbp]
    \centering
    \includegraphics[width=0.85\linewidth]{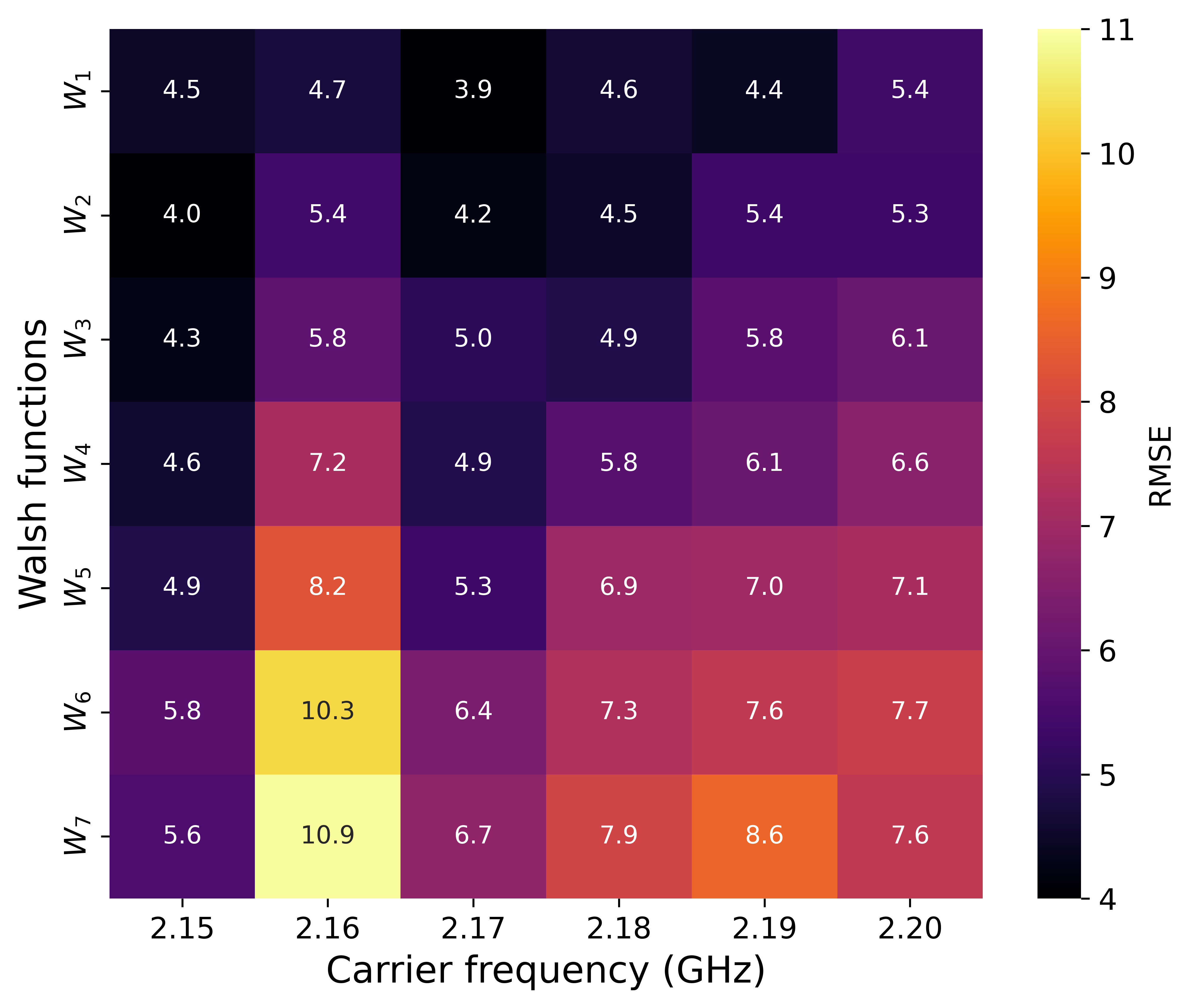}
    \caption{RMSE $\mathbf{E}$ between actual demodulated signals and their reconstruction for all Walsh functions across different drive frequencies.}
    \label{fig:E_vs_f_d_W_i}
\end{figure}
In Fig.~\ref{fig:E_vs_f_d_W_i}, $e_7$ is highest for responses corresponding to $W_7(t)$, as capturing these dynamics would require functions with faster temporal variations than $W_7(t)$ within the linear approximation.

\begin{figure}[!htbp]
\centering
\includegraphics[width=0.85\columnwidth]{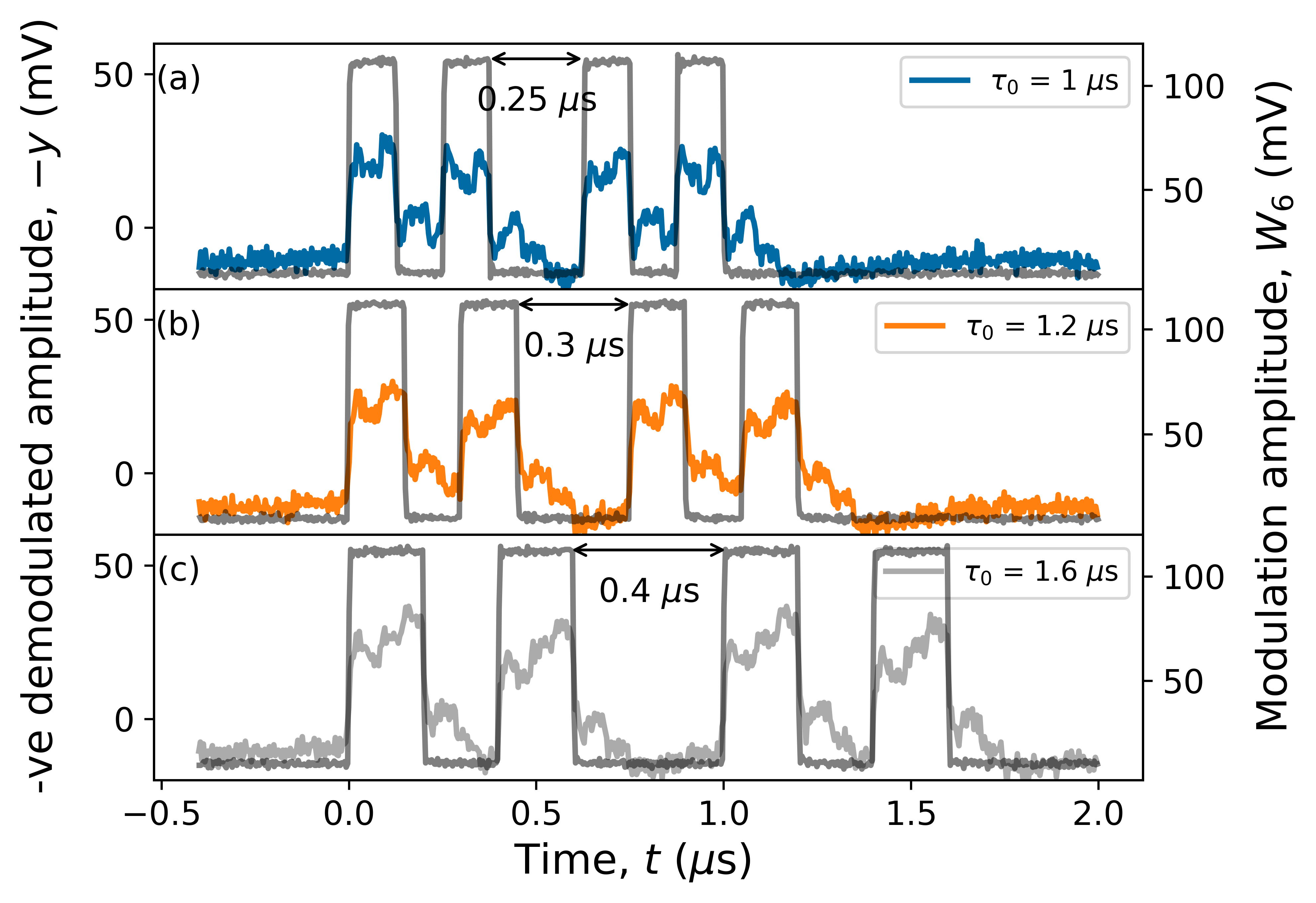}
\caption{Responses of SWARR to the drive signal modulated by $W_6(t)$, at a $f_\mathrm{d} = 2.1588~ \mathrm{GHz}$ and $G_2 = 33~\mathrm{dB}$ \textcolor{black}{for the three different code durations} $\tau_0$ = (a) 1, (b) 1.2, (c) 1.6~$\mu$s.}
\label{fig:response_w6}
\end{figure}
\begin{figure}[tbh]
\centering
\includegraphics[width=0.85\columnwidth]{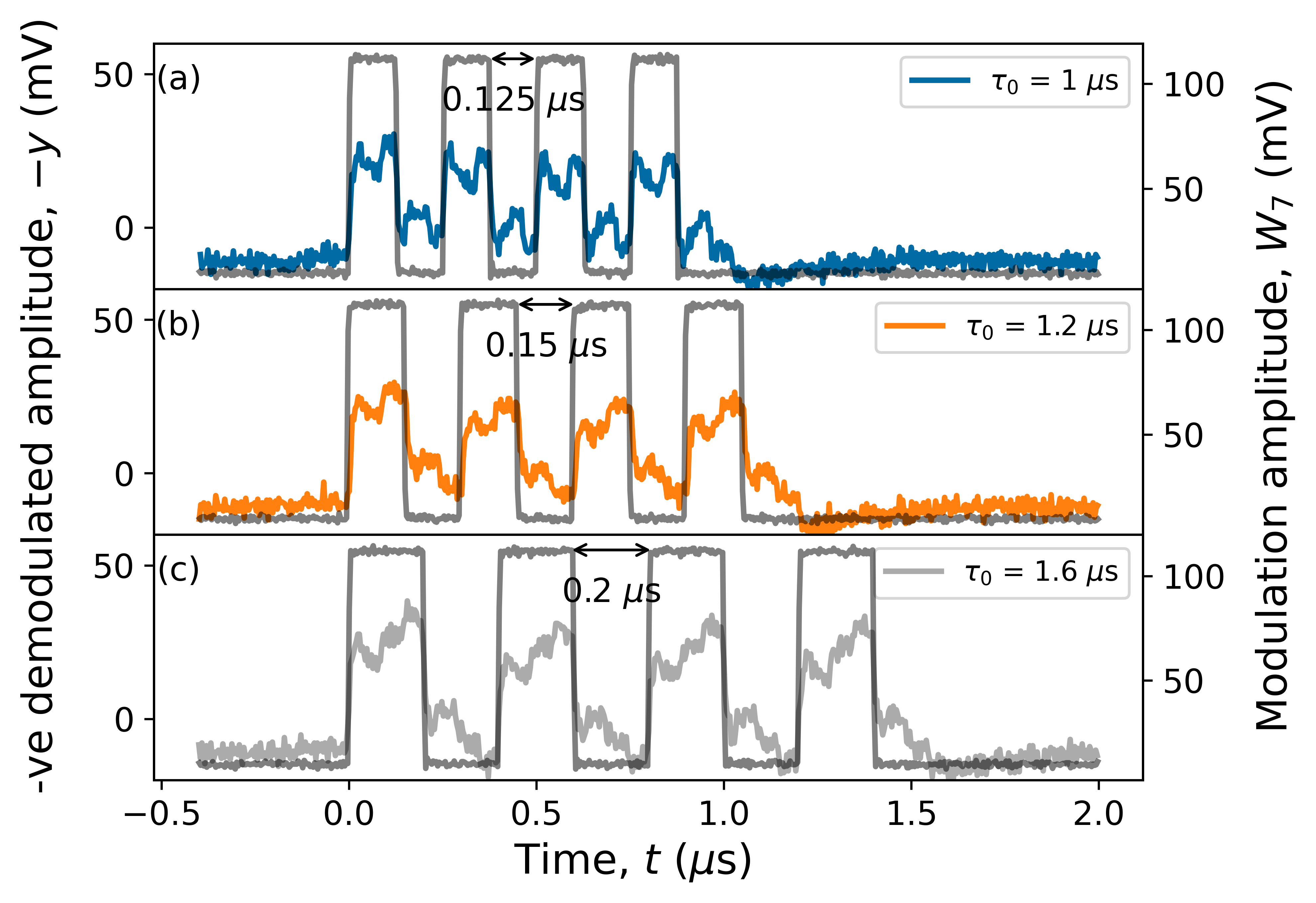}
\caption{Responses of SWARR to the drive signal modulated by $W_7(t)$, at a $f_\mathrm{d} = 2.1588~\mathrm{GHz}$ and $G_2 = 33~\mathrm{dB}$ \textcolor{black}{for the three different code durations} $\tau_0$ = (a) 1, (b) 1.2, (c) 1.6~$\mu$s.}
\label{fig:response_w7}
\end{figure}
\begin{figure}[tbh]
\centering
\includegraphics[width=0.85\columnwidth]{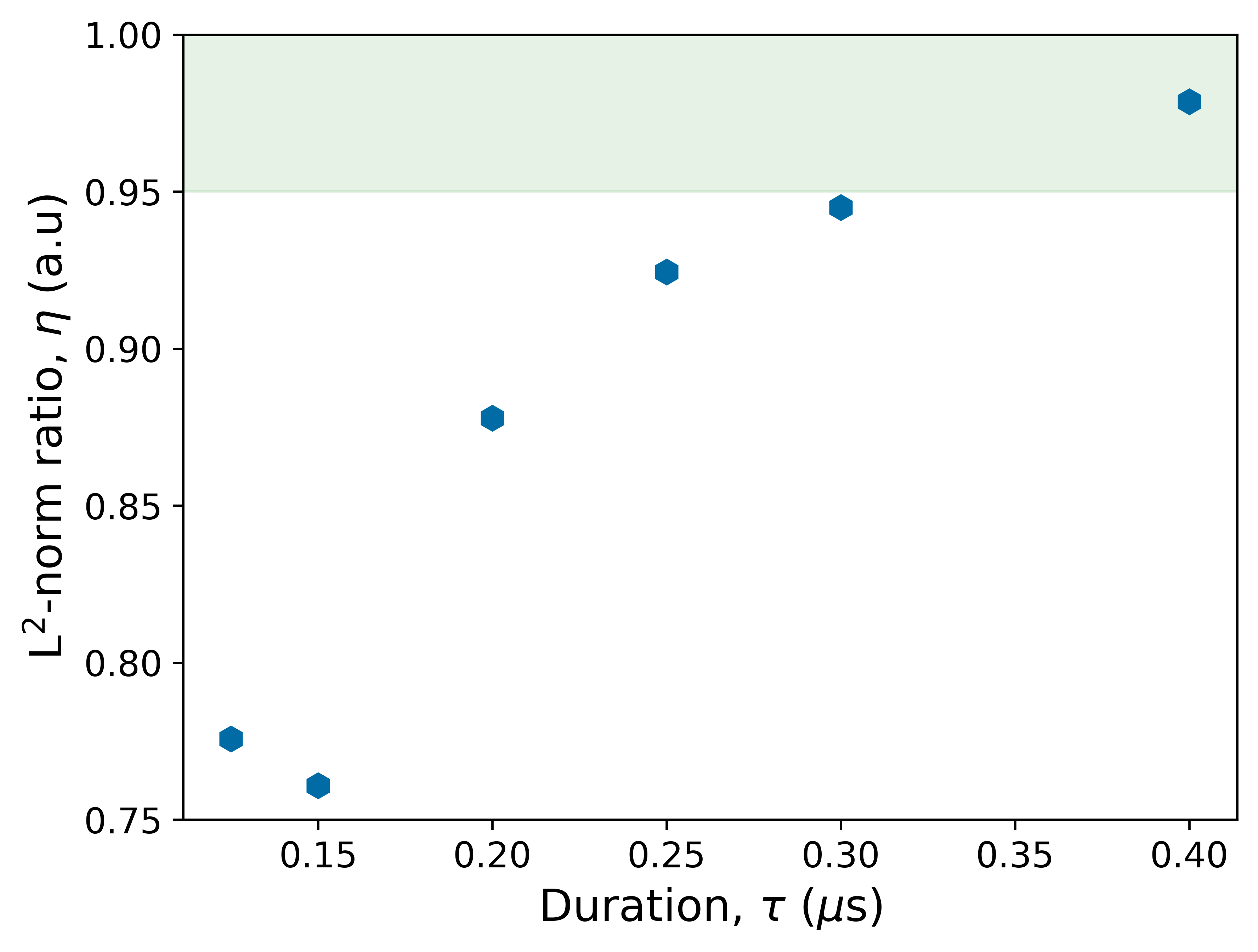}
\caption{L$^2$-norm ratio at $f_\mathrm{d} = 2.1588~\mathrm{GHz}$ and $G_2 = 33~\mathrm{dB}$, plotted as a function of $\tau$. The green region indicates the 5\% band.}
\label{fig:l2_vs_tau}
\end{figure}
The SWARR is probed using GHz carrier signals modulated by $W_{6}(t)$ and  $W_{7}(t)$ with code durations $\tau_0 = 1, 1.2$ and $1.6\,\mu$s. The STM is quantified by examining the L$^2$-norm ratio $\eta$ between the SWARR output in response to the $0^{\mathrm{th}}$ bit and the next bit with a value of 1, e.g., $b_5^{(6)}$ in the Walsh codeword $W_6$. This separation is marked as $\tau$ in Fig.~\ref{fig:w_67}.

L$^2$-norm ratio between
two signals $p(t)$ and $q(t)$ of length $N$ is defined as:
% \begin{equation}
%     R_{pq}^{\mathrm{max}} =  \frac{\max_{\ell}\left|\displaystyle\sum_{n=0}^{N-1} p[n]\, q[(n+\ell)]\right|}
%     {\sqrt{\displaystyle\sum_{n=0}^{N-1} p[n]^2 \cdot \sum_{n=0}^{N-1} q[n]^2}},
%     \quad \ell = 0, 1, \ldots, N-1,
%     \label{eq:xcorr}
% \end{equation}
% and the L$^2$-norm ratio is given as:
\begin{equation}
    \eta = \sqrt{\frac{\displaystyle \sum_{n=0}^{N-1} p[n]^2}{\displaystyle \sum_{n=0}^{N-1} q[n]^2}}
    \label{eq:l2_norm}.
\end{equation}
The demodulated SWARR responses to $W_6(t)$ and $W_7(t)$ are shown in 
Figs.~\ref{fig:response_w6} and~\ref{fig:response_w7}, respectively.
% The vertical dashed line pairs in Figs.~\ref{fig:response_w6_1} and \ref{fig:response_w7_1} mark the SWARR outputs
Let $y(b_k^{(i)})$ denote the SWARR output in response to bit 
$b_k^{(i)}$ of codeword $W_i$, where $(i,k) \in 
\{(6,0),\,(6,5),\,(7,0),\,(7,4)\}$.
The $L^2$-norm ratios corresponding to the signal pairs $y(b_0^{(6)})$, $y(b_5^{(6)})$ and $y(b_0^{(7)})$, $y(b_4^{(7)})$ are evaluated using (\ref{eq:l2_norm}) and plotted in Fig.~\ref{fig:l2_vs_tau} as a function of $\tau$. We assume that an $\eta > 0.95$ indicates the absence of STM. At $\tau = 300~\mathrm{ns}$, $\eta$ approaches the 5\% tolerance band. Therefore, the STM duration can be estimated to be approximately 300~ns.
Since the sweep over \textcolor{black}{code duration} $\tau_0$ is relatively coarse, the precise value of $\tau$ at which the STM effect vanishes cannot be 
determined. 
% Nevertheless, the results consistently indicate that the STM duration of the SWARR is approximately 300~ns.

\section{Conclusion}
In this work, a square wave pattern with a spectral null at $\frac{3}{T}$ is employed as a modulation signal to quantify the nonlinear behavior of the SWARR. By examining the power spectra from the YIG delay line, five distinct regions are \textcolor{black}{observed} over the drive frequency range of $2.15$ to $2.2~\mathrm{GHz}$, where the spectral components at $f_\mathrm{d} \pm \frac{3}{T}$ are excited by the nonlinear response of the system. A modulation pattern derived from the sequency-ordered 8-bit unipolar Walsh family was employed to estimate the STM duration of the SWARR, yielding a value of approximately 300~ns. The nonlinear behavior of the SWARR was also characterized by decomposing its temporal response to the input Walsh analog pulses in terms of the Walsh codewords 
themselves. Together, harmonic elimination and Walsh-function decomposition constitute a practical and generalizable framework for the design and optimization of tunable spin-wave reservoir computers.
%
% \appendices
% \section{}
% \begin{figure}[!htbp]
% \centering
% \includegraphics[width=0.85\columnwidth]{20260302_222425_swarr_response_to_w6_Im_1.8A_G_33dB_(1).png}
% \caption{Responses of SWARR to the drive signal modulated by $W_6(t)$, at a $f_\mathrm{d} = 2.1588~ \mathrm{GHz}$ and $G_2 = 33~\mathrm{dB}$ \textcolor{black}{for the three different code durations} $\tau_0$ = (a) 1, (b) 1.2, (c) 1.6~$\mu$s. The vertical line pairs mark the SWARR outputs corresponding to $b_0^{(6)}$ and $b_5^{(6)}$.}
% \label{fig:response_w6_1}
% \end{figure}
% %
% \begin{figure}[tbh]
% \centering
% \includegraphics[width=0.85\columnwidth]{20260302_222425_swarr_response_to_w7_Im_1.8A_G_33dB_(1).png}
% \caption{Responses of SWARR to the drive signal modulated by $W_7(t)$, at a $f_\mathrm{d} = 2.1588~\mathrm{GHz}$ and $G_2 = 33~\mathrm{dB}$ \textcolor{black}{for the three different code durations} $\tau_0$ = (a) 1, (b) 1.2, (c) 1.6~$\mu$s. The vertical line pairs mark the SWARR outputs corresponding to $b_0^{(7)}$ and $b_4^{(7)}$.}
% \label{fig:response_w7_1}
% \end{figure}
%
\section*{Acknowledgment}
The authors would like to thank N. Bilanuik and D. D. Stancil for the micro-strips. A.M. is also thankful to the Mphasis F1 Foundation for financial support.
% references section
\bibliographystyle{IEEEtran}
\bibliography{REF.bib}
\end{document}